\documentstyle[12pt,bbbold,psfig,a4]{article}

\def\slantfrac#1#2{\hbox{$\,^{#1}\!/_#2$}}

\begin{document}

\title{Orbital evolution of a binary neutron star and
gravitational radiation}
\author{Kuznetsov O.A.$^1$, Prokhorov M.E.$^2$,\\
Sazhin M.V.$^2$, Chechetkin V.M.$^1$\\[0.5cm]
$^1${\it Keldysh Institute of Applied Mathematics, Moscow}\\
{\sf kuznecov@spp.keldysh.ru; chech@int.keldysh.ru}\\[0.5cm]
$^2${\it Sternberg Astronomical Institute, Moscow}\\
{\sf mike@sai.msk.ru; sazhin@sai.msk.ru}\\[0.5cm]
{\it Astronomy Reports, Vol.42, No.5, 1998, pp.638--648}\\
{\it Translated from  Astronomicheskii Zhurnal}\\
{\it Vol.75, No.5, 1998, pp.725--735}}

\date{}

\maketitle

\begin{abstract}
{\bf Abstract}---We consider the dependence of the internal
structure of a neutron star in a close binary system on the
semi-major axis of the binary orbit, focusing on the case when the
Roche lobes of the components are nearly filled. We adopt a
polytropic equation of state. The temporal evolution of the
semi-major axis and its dependence on the mass ratio of the binary
components and the polytropic index $n$ are determined. The
calculation are carried out right up to the moment of contact,
when quasi-stationary model becomes invalid. We analyze
differences in the shapes of the pulses of gravitational radiation
emitted by a pair of point masses and by a binary neutron star,
taking into account its internal structure and tidal deformations.
\end{abstract}

\section{INTRODUCTION}

Gravitational-wave astronomy is currently one of the most
intensively developing branches of science. Coalescing neutron
stars are especially interesting sources of gravitational
radiation in the context of future gravitational-wave detectors.
Binary neutron star exist in our Galaxy, and are observed as
binary radio pulsars. Such systems lose energy via gravitational
radiation; this causes the semi-major axis of the system orbit
to decrease, ultimately leading to the coalescence (merging) of
the stars. A large number of such systems for which the merging
time is much less than the cosmological time are known, so that
neutron stars merges should occur fairly often (no more rarely
than about 3 events per year within a distance of 200 Mpc)
[1--4]. Already operating and planned laser gravitational-wave
detectors [3, 5, 6] will search for such events.

The merging process can be divided into three physically
distinct stages.

\begin{enumerate}
\item The binary components move in quasi--Keplerian orbits
without mass transfer, approaching each other only due to the
emission of gravitational radiation.
\item Upon the stars become sufficiently close, mass transfer
between the binary components must be taken into account in
addition to the Keplerian motion and gravitational radiation.
\item In a late stage of the mass transfer, the merging of the
binary components or falling of on star onto the other leads to
the formation of a single object.  This is the third stage of
merging process. (Note that the distinction between the second
and the third stages is quite arbitrary.)
\end{enumerate}

This problem has been treated in a large number of studies,
among which two different approaches can be distinguished. The
first approach is analytical. It is useful for studies of very
distant pairs, in which distortions in the shape of the stars
due to their mutual gravitational attraction and other tidal
effects are negligible. In this case, the problem can be reduced
to a form that permits finite, analytical solution. A number of
such studies account for post--Newtonian gravitational effects
[7--9]. This approach fails for even moderately close pairs, in
which tidal effects, rotation, and the internal structure of the
components become substantial, and must be taken into account
when deriving information from the gravitational-wave pulses
emitted [10].

The second approach relies on numerical computations. It uses
hydrodynamic codes to describe the non-stationary coalescing
and merging of two bodies, taking into account a whole series of
physical processes that may be important. This approach is the
only able to describe the late stages of a merger [11].
Unfortunately, current hydrodynamic codes have a number of
drawbacks, including the presence of artificial viscosity in the
numerical algorithms used and the inability to follow the
evolution of a binary system from large values of the semi-major
axis (say, $\sim 10^6$ stellar radii) up to contact.

The first (analytical) method is able to trace the binary
evolution up to a hundred stellar radii rather accurately, while
the second (hydrodynamical) method can be used to calculate the
evolution of the system beginning from several stellar radii.
Thus, there remains an appreciable interval of semi-major axis
values (say, from 100 to 5--10 $R_\star$) for which different
methods must be applied. We describe below one possible method
in which quasi-stationary stellar configurations are calculated
fully taking into account tidal effects.

We will focus on a detailed description of the first stage, when
the stars move only under the action of their mutual
gravitational attraction and gravitational radiation. The
problem will be solved numerically using the code developed in
[12]. The second part of the problem is the determination of the
shapes of gravitational-wave pulses emitted and identification
of characteristic features that can be used to derive the main
parameters of neutron stars.

\section{A MODEL FOR AN EQUILIBRIUM POLYTROPIC STAR IN A BINARY
SYSTEM}

We consider a model for a binary star with component masses
$M_1$ and $M_2$ and total mass $M=M_1+M_2$. We assume the orbit
to be circular with radius $A$; this assumption should be
reasonable, since the orbit should be circularize long before
the final stage of coalescence due to emission of gravitational
radiation [7].  The primary component is treated as a polytropic
star with polytropic index $n$, while the secondary is taken to
be a point mass. Below, we will present arguments showing that
this assumption is justified. As in [12], we chose the
coordinate origin to be at the center of the first component,
and the orthogonal coordinate axes to be oriented so that $Ox$
passes through the centers of the two stars and $Oy$ lies in the
orbital plane.

The equation of state of the first star is

$$
    P = K\rho^{1+1/n}
$$
where $P$ and $\rho$ are the pressure and density of the stellar
matter and $K$ is the polytropic constant. The density and the
potential distributions are described by the equation of
hydrostatic equilibrium:

$$
\Phi_1+\Phi_2+\Phi_c+K(n+1)\rho^{1/n}=B\,,\eqno(1)
$$
where $\Phi_1$ is the potential of the first star (both inside
and outside), $\Phi_2$ is the potential of the second star (point
mass)

$$
\Phi_2 = -\frac{GM_2}{\sqrt{(x-A)^2+y^2+z^2}}\,,
$$
$\Phi_c$ is the centrifugal potential, and $B$ is a constant
determined by the boundary conditions. The centrifugal potential
is taken in the form

$$
\Phi_c=-\slantfrac{1}{2}\Omega^2[(x-X_c)^2 +y^2]\,,
$$
where $X_c$ is the center of mass of the system. In contrast to
[12], we do not use the Keplerian formula for the semi-major
axis for a given angular velocity $\Omega$. Instead, we
determine this value self-consistently, taking into account the
tidal interaction of the stars:

$$
\frac{\partial}{\partial
x}\left.(\Phi_1+\Phi_c)\right|_{(A,0,0)}=0\,.
$$

On the other hand, the density and potential of the first star
are related by the Poisson equation

$$
\Delta\Phi_1 = 4\pi G\rho\,.\eqno(2)
$$

To find the equilibrium configuration for the extended component
of the system, we must solve the system of equation (1) and (2)
using the condition

$$
\int \rho_1 dV_1=M_1\,.
$$

To implement the numerical algorithm, it is convenient to use
the following dimensionless variables. We will express distances
in units $L=(GM/\Omega^2)^{1/3}$, masses in units
of $M_2$, densities in units of $M_2/L^3$, the gravitational
potential in units of $GM_2/L$, and the polytropic constant in
units of $GM_2^{1-1/n}L^{3/n-1}$. As a result, we arrive at the
dimensionless equations

$$
\phi_1+\phi_2+\phi_c+\kappa (n+1)\rho^{1/n}=b\,,\eqno(3)
$$

$$
\Delta\phi_1 = 4\pi \rho\,,\eqno(4)
$$

$$
\int\rho dV = q\,,\qquad q=M_2/M_1\,,\eqno(5)
$$

$$
a=A/L\,,\eqno(6)
$$

$$
\phi_2 = -\frac{1}{\sqrt{(x-a)^2+y^2+z^2}}\,,\eqno(7)
$$

$$
x_c=q^{-1}\int x\rho dV=0\,,\eqno(8)
$$

$$
\phi_c=-\slantfrac{1}{2}(q+1)\left(\left(x-\frac{a}{1+q}
\right)^2+y^2\right)\,,\eqno(9)
$$

$$
\frac{\partial}{\partial
x}\left.(\phi_1+\phi_c)\right|_{(a,0,0)}=0\,,\eqno(10)
$$

$$
\phi_1 \to -q/\sqrt{x^2+y^2+z^2}\,,\qquad x,y,z \to
\infty\eqno(11)
$$
in which lower-case letters denote dimensionless variables,
$\kappa$ is the dimensionless polytropic constant, and $\rho$ is
now the dimensionless density.

\section{NUMERICAL ALGORITHM}

We transformed the system of equations to dimensionless form and
solved it using the iterative method described in detail in
[12]. We will now briefly summarize this method.

Let $M_1$, $M_2$, $\Omega$ $K$, and $n$, be given, while $A$,
$B$, $\rho(x,y,z)$, and $\Phi_1(x,y,z)$ must be determined. We
do not assume {\it a priori} that Kepler's third law
$A^3\Omega^2=GM$ is satisfied. In dimensionless form,
the input parameters are $q$, $\kappa$, and $n$ (note that
$m_2=1$, $m_1=q$, and $\omega=\sqrt{q+1}$ in dimensionless
form), while $a$, $b$, $\rho(x,y,z)$, and $\phi_1(x,y,z)$ must be
found. The dimensionless form of Kepler's third law is $a=1$.
All calculations were performed for the case $q=1$ and
$n=\slantfrac{3}{2}$, and $\kappa$ was varied. This is in
accordance with the relation $K=\kappa GM_2^{1-1/n}L^{3/n-1}$,
which is equivalent to the shrinking of the system with fixed
polytropic constant $K$, with $\kappa\to0$ corresponding to an
infinitely wide system ($L\to\infty$).

The calculations can be applied to systems with any $M_1$ and
$K$. However, to make the results more transparent, we will scale
them to correspond to a binary system in which
$M_1=M_2=1.4M_\odot$ and the radius of unperturbed neutron star
is 10 km. In this case of an infinitely wide pair, when the
first component is purely spherical, this corresponds to an
average density for the first component
${\bar\rho}_1=6.68\times10^{14}{\rm g/sm}^3$ and a polytropic
constant $K=3.99\times10^{9}{\rm cm}^4{\rm g}^{-2/3}{\rm
s}^{-2}$.

The scheme for the solution of the system of equation (3)--(11)
is, in many ways, analogous to that in [12], with the addition
of two more iteration cycles to solve the equations $x_c=0$ and
(10). These differences from [12] are associated with the
necessity of determining $a$, which, in general, differs from
unity (i.e. Kepler's third law is not necessarily valid for a
distributed mass). We note, however, that the value for $a$
returned by the computer code deviates from the unity by less
then 0.5\%.

\section{GRAVITATIONAL RADIATION FROM A BINARY SYSTEM WITH AN
EXTENDED COMPONENT}

Energy losses due to emission of gravitational waves are
described by the formula [13]

$$
-\frac{\partial E}{\partial t}=\frac{G}{45c^5}
\stackrel{\bullet\bullet\bullet}{Q}_{\alpha\beta}^2\,,\eqno(12)
$$
where

$$
Q_{\alpha\beta}=
\int\rho
\left(x_\alpha x_\beta-\slantfrac{1}{3}\delta_{\alpha\beta}x^2
\right)dV
$$

Since the rate of change of the orbit at the end of its
evolution is comparable to the velocities of the components, we
must, in general, take into account the first and higher time
derivatives of the semi-major axis.

The main formulas for our analysis are presented for a rotating
coordinate system. For transformation of quantities to a
non-rotating frame, we write expressions for quadrupole moments
of the system and of individual bodies in non-rotating and
rotating coordinate systems. We denote the non-rotating frame by
$O{\hat x}{\hat y}{\hat z}$, and add a sign $~\hat{}~$
over all quantities related to this frame. We keep the notation
$Oxyz$ for the rotating frame, and quantities in this frame will
be presented without a sign $~\hat{}~$. The
transformations from the rotating to the non-rotating coordinate
system are given by equations

$$
x=\hat x \cos \psi(t) -\hat y \sin \psi(t)\,,
$$

$$
y=\hat x \sin \psi(t) +\hat y \cos \psi(t)\,,
$$
where $\stackrel{\bullet}{\psi}(t)\equiv\Omega(t)$ is the
time-dependent angular velocity of the binary system.

Recall that our model for the binary system consists of a
primary component that is a polytropic star under hydrostatic
equilibrium in the rotating frame and a secondary component that
is a point mass. The quadrupole tensor of such a binary system
in the rotating frame can be written as

$$
Q_{\alpha\beta}=m_1(r_{1\alpha}r_{1\beta}-{\slantfrac{1}{3}}\delta_{\alpha
\beta}r_1^2)+m_2(r_{2\alpha}r_{2\beta}-{\slantfrac{1}{3}}\delta_{\alpha
\beta}r_2^2)+q_{\alpha \beta}\,,
$$
where we have introduced the notation

$$
q_{\alpha \beta}=\int d^3\xi\rho_1(\xi)(\xi_{\alpha}\xi_{\beta}
-{\slantfrac{1}{3}}\delta_{\alpha\beta}\xi^2)
$$
for the quadrupole tensor of the distributed mass, while the
first and second terms are the contributions to the
quadrupole tensor of two point masses located at ${\bmath
r}={\bmath r}_1$ and ${\bmath r}={\bmath r}_2$ respectively.
Here, $\xi$ are the coordinates relative to the center of mass
of the first star ${\bmath r}={\bmath r}_1$.

It is known [13] that the quadrupole tensor can be expressed in
terms of the inertia tensor of the first star

$$
I_{\alpha\beta}=\int d^3\xi\rho_1(\xi)(\delta_{\alpha\beta}\xi^2
-\xi_{\alpha}\xi_{\beta})
$$
in accordance with the expressions

$$
q_{\alpha \beta}={\slantfrac{1}{3}}\delta_{\alpha\beta}I-I_{\alpha\beta}\,,
$$
where $I=I_{XX}+I_{YY}+I_{ZZ}$ is the trace of the inertia
tensor.

Since ${\bmath r}$ is defined for the baricenter of the system,
we can find explicit expressions for the vectors of each
component in the form

$$
{\bmath r}_1=\frac{m_2 A}{m_1+m_2}(1,\; 0,\; 0)\,,
$$

$$
{\bmath r}_2=\frac{m_1 A}{m_1 +m_2}(-1,\; 0,\; 0)\,.
$$

Now, the explicit form of the quadrupole tensor in the rotating
frame will be a function of two parameters: the quadrupole tensor
of the system of point masses, which is determined by $\mu A^2$
($\mu$ is the reduced mass of the binary system), and the
quadrupole tensor of the extended component
$q_{\alpha\beta}$. Since our model is symmetrical with respect
to the planes $y=0$ and $z=0$, only three of nine components of
the quadrupole tensor are non-zero: $q_{XX}$, $q_{YY}$, and
$q_{ZZ}$. Since the trace is equal to zero
($q_{XX}+q_{YY}+q_{ZZ}=0$), there remain only two independent
parameters, which fully describe the total quadrupole tensor of
the binary system in the rotating frame:

$$
f_1={\mu A^2 \over 6} +\frac{q_{XX} +q_{YY}}{2}\,,
$$

$$
f_2={\mu A^2 \over 2} +\frac{q_{XX} -q_{YY}}{2}\,.
$$

We now write the components of the quadrupole tensor in the
non-rotating frame:

$$
\hat Q_{XX}=f_1 +f_2 \cos 2\psi (t)\,,
$$

$$
\hat Q_{XY}= -f_2 \sin 2\psi (t)\,,
$$

$$
\hat Q_{YY}=f_1 -f_2 \cos 2\psi (t)\,,
$$

$$
\hat Q_{ZZ}= -2f_1\,,
$$

$$
\hat Q_{XZ}= \hat Q_{YZ}=0\,.
$$

In the case of stationary orbit
($\stackrel{\bullet}{\Omega}=0$), the formula for gravitational
radiation differs from the standard formula of two point masses
[13] only in a term of the form $q_{XX}-q_{YY}$:

$$
-\frac{d E}{d t}=\frac{32G}{5c^5}\Omega^6(\mu
A^2+q_{XX}-q_{YY})^2
$$
This formula reduces to the known relation [13]

$$
-\frac{d E}{d t}=
\frac{32G}{5c^5}\Omega^6(\mu A^2)^2=
\frac{32G}{5c^5}\Omega^6A^4\left(\frac{M_1M_2}{M}\right)^2
$$
in the case of point masses ($q_{XX}=q_{YY}=0$). Given the
density distribution in the extended component of the system,
we can calculate the dimensionless quadrupole momentum of this
component with respect to the center of mass. Let us introduce
the dimensionless function

$$
f=\frac{q}{1+q}\left[\left(\frac{A}{L}\right)^2-1\right]
+\frac{q_{XX}-q_{YY}}{M_2L^2}
$$
so that the luminosity of the binary system in the form of
gravitational waves is

$$
-\frac{d E}{d t}= \frac{32G^4 M_2^5}{5c^5L^5}
\left(\frac{q}{1+q}+f\right)^2 (1+q)^3
$$

Using the relation $K=\kappa GM_2^{1-1/n}L^{3/n-1}$, we
eliminate $L$ and obtain

$$
-\frac{d E}{d t}=
\frac{32G^{\frac{12+n}{3-n}}M_2^{\frac{10}{3-n}}}{5c^5K^{\frac{5n}{3-n}}}
\kappa^{\frac{5n}{3-n}}
\left(\frac{q}{1+q}+f(\kappa)\right)^2(1+q)^3
$$

Note that, in the case of binary systems in which extended
components are elongated along the line through the stellar
centers, taking into account the proper moment of inertia of a
component will always lead to an increased energy-loss rate
compared to the case of point masses.

\section{TOTAL ENERGY OF THE SYSTEM IN A MODEL WITH AN EXTENDED
COMPONENT}

If the energy loss $\stackrel{\bullet}{E}$ is known, we can
calculate the shrinking velocity of the stars
$\stackrel{\bullet}{E}=dE/dA\,\cdot\stackrel{\bullet}{A}$. In the
standard approach [13], the energy of the system is taken to be
the sum of gravitational energy of two point passes $E_{12}$ and
their kinetic energies $K_1$ and $K_2$. This yields an
expression for the energy of the system $E=-GM_1M_2/2A$. If the
masses are not point-like, we must also take into account the
self-gravitational energies $E_1^{self}$ and $E_2^{self}$, the
internal energies $E_1^{int}$ and $E_2^{int}$, and the kinetic
energies of rotation of the stars about their centers of mass
$K_1^{*}$ and $K_2^{*}$. We will take the second component to be
point so that $E_2^{self}=E_2^{int}=K_2^{*}=0$. Then

$$
E=E_{12}+K_1+K_2+E_1^{self}+E_1^{int}+K_1^{*}\,,
$$

$$
E_{12}=\int\rho\Phi_2dV_1\,,\qquad
E^{self}_1=\slantfrac{1}{2}\int\rho\Phi_1dV_1\,,
$$

$$
K_1=\slantfrac{1}{2}\Omega^2M_1\left(\frac{M_2}{M}A\right)^2\,,
\qquad
K_2=\slantfrac{1}{2}\Omega^2M_2\left(\frac{M_1}{M}A\right)^2\,,
$$

$$
K^*_1=\slantfrac{1}{2}\Omega^2I_{ZZ}\,,\qquad
E^{int}_1=n\int PdV_1=nK\int\rho^{1+1/n}dV_1\,.
$$

Let us now consider the asymptotic of $E^{int}_1$, $E^{self}_1$
as $L\to\infty$. In this case, the first component can be
considered a polytropic, self-gravitating sphere. The solution
of the equilibrium equations for such configurations are well
known [14,15]. The total energy of a polytropic,
self-gravitating sphere in hydrostatic equilibrium is given by
the formula [14,16]

$$
E^{self}_1+E^{int}_1=-\frac{3-n}{5-n}\frac{GM^2_1}{R_1}\,.
$$

The polytropic constant of such a sphere is determined by its
mass and radius [14]:

$$
K={\cal N}_nGM_1^{\frac{n-1}{n}}R_1^{\frac{3-n}{n}}\,.
$$

The coefficient ${\cal N}_n$ is associated with the solution of
Lane-Emden equation\footnote{The Lane-Emden equation has the
form

$$
\xi^{-2}(\xi^2\theta')'=-\theta^n\,,
$$
where $\xi$ is the independent variable and $\theta(\xi)$ is the
desired function ($\theta'\equiv d\theta/d\xi$), with the
boundary conditions given by

$$
\theta(0)=1\,,\qquad \theta'(0)=0\,,
$$
and by the values at the boundary of the sphere $\xi_1$:
$\theta(\xi_1)=0$, and $\mu_1=-\xi_1^2\theta'(\xi_1)$.}
via the formula

$$
{\cal N}_n=
(4\pi)^{1/n}(n+1)^{-1}\xi_1^{-\frac{3-n}{n}}\mu_1^{-\frac{n-1}{n}}\,,
$$

$$
{\cal N}_{3/2}=0.424\,.
$$

On the other hand,

$$
K=\kappa GM_2^{\frac{n-1}{n}}L^{\frac{3-n}{n}}\,.
$$
We find as a result

$$
E^{self}_1+E^{int}_1 \to E_0
\left(-\frac{3-n}{5-n}{\cal N}_n^{\frac{n}{3-n}}q^{\frac{5-n}{3-n}}
\right)=E_\infty\,, \qquad L \to \infty\,,
$$

$$
E_0=G^{\frac{3}{3-n}}M_2^{\frac{5-n}{3-n}}K^{-\frac{n}{3-n}}\,.
$$
The sum of energies $E_{12}+K_1+K_2$ has the following asymptotic
as $L\to\infty$:

$$
E_{12}+K_1+K_2 =E_0
\left(-\slantfrac{1}{2}q\kappa^{\frac{n}{3-n}}\right)\,.
$$

We finally obtain for the total energy of the system the formula

$$
E = E_0
\left(-\slantfrac{1}{2}q\kappa^{\frac{n}{3-n}}
+h(\kappa)\right)+E_\infty\,.
$$
Here, $h(\kappa)$ is determined from calculations of this
function (equal to the sum $K_1^*$ and the differences of
$E^{self}_1+E^{int}_1$ and $E_{12}+K_1+K_2$ from their limiting
values), with $h(\kappa)\to 0$ as $\kappa\to0$ (i.e., as
$L\to\infty$):

$$
h(\kappa)=\frac{E-E_0\left(-\slantfrac{1}{2}q\kappa^\frac{n}{3-n}\right)
-E_\infty}{E_0}\,.
$$
Since the first component expands as the two components
become closer, \linebreak $h(\kappa)>0$.

\section{METRIC OF THE GRAVITATIONAL-RADI\-ATION FIELD}

The physical observed quantities in gravitational-wave astronomy
are the transversely traceless corrections to the metric tensor
$h_+^{TT}$ and $h_\times^{TT}$. These are expressed in terms of
time derivatives of the components of the quadrupole tensor of
the system. To write the explicit form of the corrections to the
metric tensor, we introduce a coordinate system that is rotated
by the angle $\theta$ about the $Oy$ axis. We will calculate the
transversely traceless corrections to the metric along the
$O{\hat x}$ axis in this new frame. The projection operator used
to reduce the tensor to a transversely traceless for can be
written
$$
\hat P=\left( \begin{array}{ccc} 0&0&0\\ 0&1&0\\ 0&0&1
\end{array}\right)\,.
$$
We can now obtain the transversely traceless components of the
quadrupole tensor in the new coordinate system:

$$
Q_+^{TT}=\slantfrac{3}{2}f_1\cos^2\theta
-f_2\frac{1+\sin^2\theta}{2}\cos2\psi\,,
$$

$$
Q_{\times}^{TT}=f_2\sin\theta\sin2\psi\,.
$$

The transversely traceless corrections to the metric are found
using the time derivatives of the reduced components of the
quadrupole tensor

$$
h_+^{TT}=-\frac{2G}{3c^4 R_0}\stackrel{\bullet\bullet}{Q}_+^{TT}
$$

$$
h_{\times}^{TT}=-\frac{2G}{3c^4
R_0}\stackrel{\bullet\bullet}{Q}_{\times}^{TT}
$$
where $R_0$ is the distance from the observer to the binary
system.

\section{EQUATION FOR THE CONVERGENCE OF THE COMPONENTS;
CALCULATION RESULTS}

The equation describing the convergence of a point mass and an
extended component differs from that for two point masses. We
will now determine how these equations change in our model and
find the dependence of the main characteristics of the
radiation on the parameters of the problem.

\begin{figure}[t]
\hbox{\psfig{figure=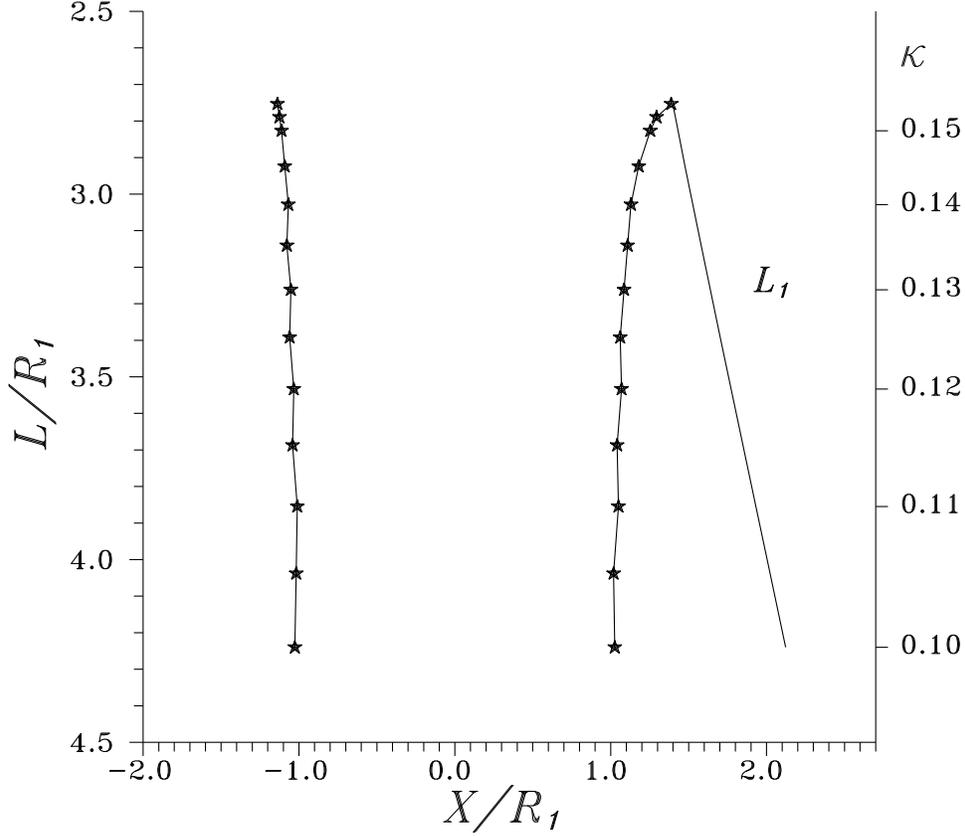,width=5in}}
\caption{\sl Geometrical characteristic of the system. The $Ox$
axis shows the boundary of the extended component: the left mark
corresponds to the far side with respect to $L_1$, and the right
mark to the near side. $R_1$ is the radius of the extended
component for an infinitely distant point-mass component. The
solid line shows the position of the Lagrange point $L_1$. The
dimensionless polytropic constant $\kappa$ is indicated on the
supplementary vertical axis. The contact of the extended
component with $L_1$ occurs at $L=2.75 R_1$.}
\end{figure}

\begin{figure}
\hbox{\psfig{figure=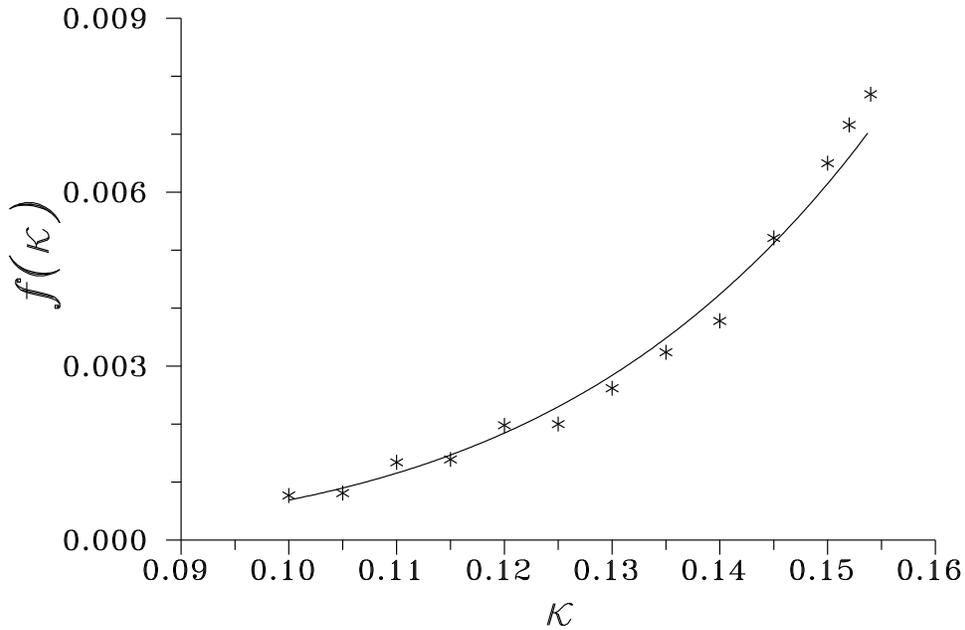,width=5in}}
\caption{\sl Plot of the function $f(\kappa)$ showing how much
the intensities of gravitational-wave radiation differ for
extended-component and point-mass models. The solid line is the
approximation $f(\kappa) \approx 171\kappa^{5.4}$.}
\end{figure}

\begin{figure}
\hbox{\psfig{figure=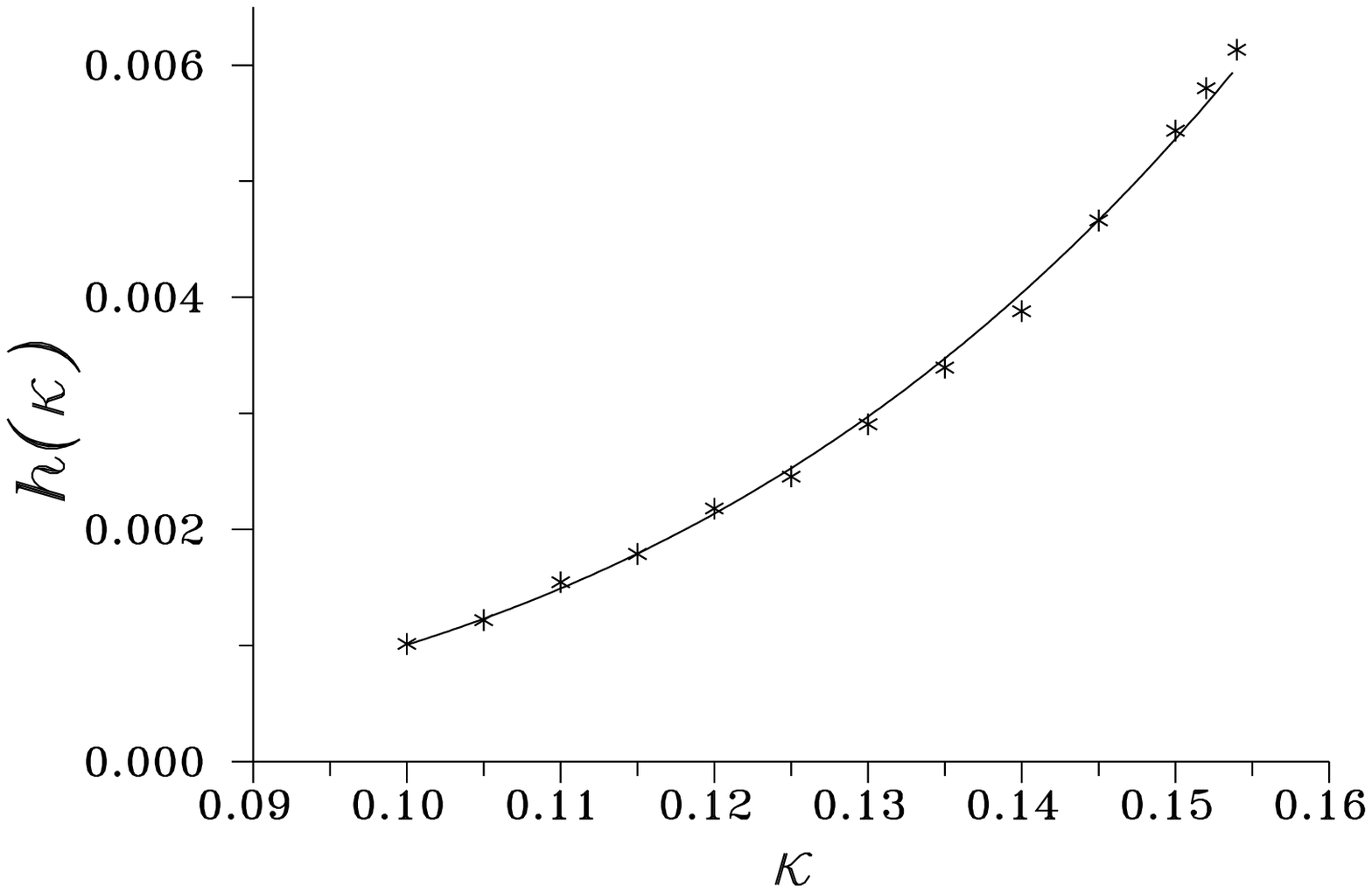,width=5in}}
\caption{\sl Plot of the function $h(\kappa)$ showing how the
total energy of a system with an extended component changes as
to the total energy for two point masses. The solid line is the
approximation $h(\kappa) \approx 13.5\kappa^{4.1}$.}
\end{figure}

First, some introductory words are necessary. The semi-major
axis of the system is denoted $A$, while the main calculations
use the arguments $L$. We emphasize that the difference between
$L$ and $A$ is small. Therefore, we will write all functions as
arguments of $L$, and assume $A$ and $L$ to be identical. The
geometrical properties if the system for various $\kappa$ are
depicted in Fig.~1, which shows the boundary of the star along
the $X$-axis (asterisks) and the position of the point $L_1$.
The horizontal and vertical axes plot $X$ and $L_1$, both
measured in terms of $R_1$. In addition, the values if
dimensionless polytropic constant $\kappa$ are marked on the
axis to the right. We can see from Fig.~1 that the components
are in contact if $\kappa=0.153$, which is corresponds to
$L/R_1=2.75$ (these estimates are independent of the chosen
values of $M_1$ and $K$, and are entirely determined by $q$ and
$n$). In addition, we can see from Fig.~1 that the size of the
first component is very insensitive to the location of the
second component when $L/R_1>3.5$: $X_{min}=X_{max}=R_1$.

Since the resulting formulas are quite cumbersome, we will
restrict the following discussion to the case
$n=\slantfrac{3}{2}$, $q=1$, for which we performed our
calculations. We then have

$$
-\frac{d E}{d t}=
\frac{256G^9M_2^{20/3}}{5c^5K^5}
\kappa^5
\left(\slantfrac{1}{2}+f(\kappa)\right)^2\,.
$$

A plot of the function $f(\kappa)$ is given in Fig.~2, in which
the solid line shows an approximation to this function using the
formula $f(\kappa)=171\kappa^{5.4}$. We then have for the total
energy of the system

$$
E = \frac{G^2M_2^{7/3}}{K}
\left(-\frac{\kappa}{2}-\slantfrac{3}{7}{\cal
N}_{3/2}+h(\kappa)\right)\,.
$$

The function $h(\kappa)$ was also calculated. This function can
be approximated by the relation $h(\kappa)=13.5\kappa^{4.1}$
(Fig.~3).

We performed the calculations for a system consisting of one
point mass and one extended component (this was associated with
limitations of our computational capabilities, and this
restriction will be relieved in the future). The results can be
applied to two types of astrophysical objects:

\begin{itemize}
\item binary systems consisting of a black hole and neutron star
of the same masses;
\item binary systems consisting of two identical neutron stars.
\end{itemize}

The first case seems unrealistic because of evolutionary
considerations, though there do exist models that predicts the
formation of low-mass black holes in binaries [17].

\begin{figure}[t]
\hbox{\psfig{figure=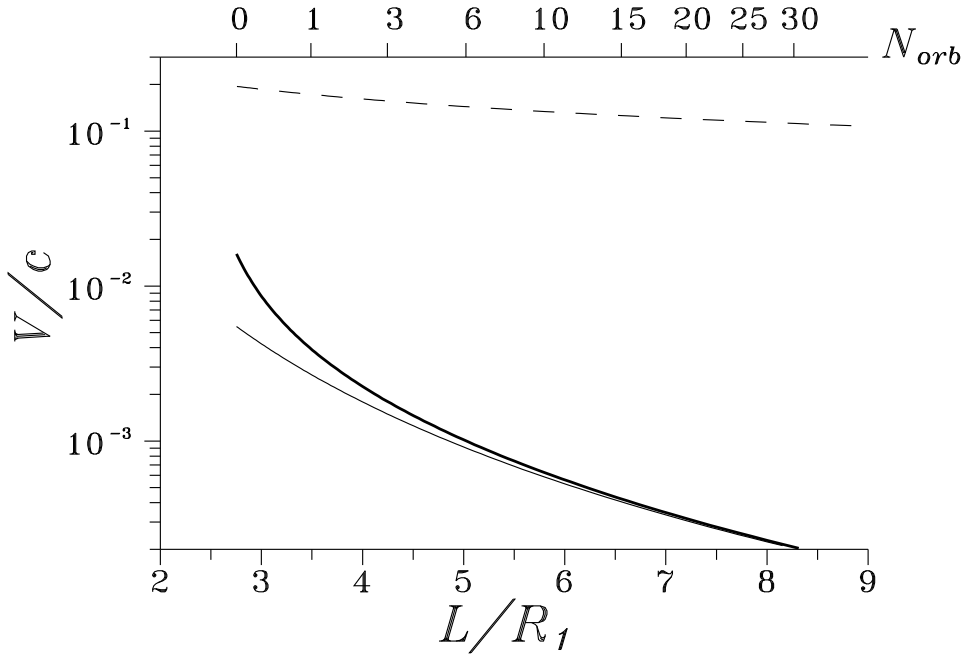,width=5in}}
\caption{\sl Velocity of one of the component with respect to the
center of mass of the system. The thick and thin lines
correspond to the cases of extended and point masses,
respectively. The dashed line shows the orbital velocity of the
system $\slantfrac{1}{2}A\Omega$.}
\end{figure}

\begin{figure}[t]
\hbox{\psfig{figure=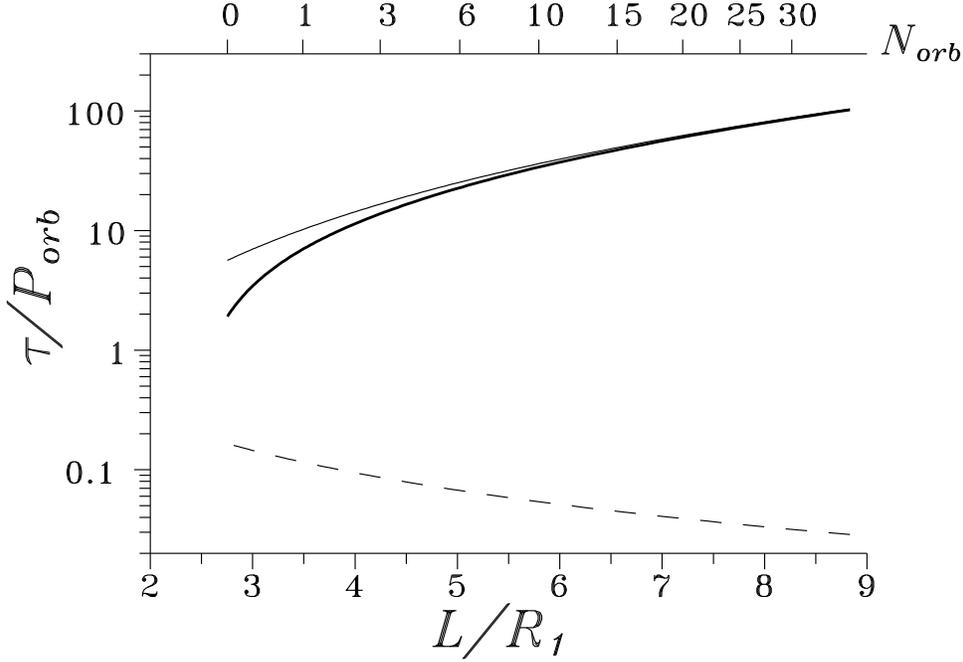,width=5in}}
\caption{\sl Rate of radial shrinkage of the components
$\tau=L/\stackrel{\bullet}{L}$ in units of the orbital period.
The thick and thin lines correspond to the case of extended and
point masses, respectively. The dashed line shows the
characteristic hydrodynamic time for the extended component,
which indicates how quickly equilibrium is reached inside the
star. The number of orbital periods until contact is indicated
on the upper axis.}
\end{figure}

We believe the second case to be more interesting. Using the
above results, we can study such a system quite easily. Suppose
the secondary component is also extended, and has the same
configuration as the primary component, since their masses are
equal. In this case the change associated with substituting the
secondary point mass by an extended mass distribution is a
correction of the next highest order, which can be neglected. It
is necessary to double the corresponding terms in the formulas
for the energy loss rate and the total energy of the system
($f(\kappa)$, $h(\kappa)$, and the term corresponding
$E_\infty$):

$$
-\frac{d E}{d t}=
\frac{256G^9M_2^{20/3}}{5c^5K^5}
\kappa^5
\left(\slantfrac{1}{2}+2f(\kappa)\right)^2\,,
$$

$$
E = \frac{G^2M_2^{7/3}}{K}
\left(-\frac{\kappa}{2}-2\slantfrac{3}{7}{\cal
N}_{3/2}+2h(\kappa)\right)\,.
$$

Differentiating the second formula, we obtain

$$
\frac{d E}{d t}=\frac{G^2M_2^{7/3}}{K}
\left(-\slantfrac{1}{2}+2h'(\kappa)\right)\frac{d\kappa}{d t}
$$
and finally

$$
\frac{d \kappa}{d t}
=\frac{256G^7M_2^{13/3}}{5c^5K^4}
\frac{\kappa^5\left(\slantfrac{1}{2}+2f(\kappa)\right)^2}
{\slantfrac{1}{2}-2h'(\kappa)}\,.
$$
Using the relations

$$
\kappa=\frac{K}{GM_2^{1/3}L}\,,
$$

$$
\frac{d L}{d t}
=-\frac{GM_2^{1/3}L^2}{K}
\frac{d \kappa}{d t}\,,
$$
we can now obtain a differential equation for $L$:

$$
-\frac{d L}{cd t}=\slantfrac{16}{5}\left(\frac{r_g}{L}\right)^3
\alpha(L)\,, \eqno(13)
$$
where $r_g=2GM_2/c^2$, which is $r_g=4.15$~km for the adopted
value of $M_2$. The auxiliary quantity $\alpha$ is defined by the
relation:

$$
\alpha(L)=1+15.5\cdot\left(\frac{R_1}{L}\right)^{3.1}+13.3
\cdot\left(\frac{R_1}{L}\right)^{5.4}\,,\eqno(14)
$$
where $R_1=10$ km is the radius of the unperturbed neutron star.
Note that the maximum value $\alpha=1.73$ is reached at the
moment of contact of the components, when $L=2.75R_1$. In the
beginning of our calculations $L=8.5R_1$, $\alpha=1.02$. The
accuracy of this formula is better than 1\% (this maximum value
is reached at the point of contact). It is not difficult to
verify that our results returns the standard expression

$$
\frac{d L}{d t} =-\frac{128G^3M_2^3}{5c^5L^3}
$$
for a point-mass binary when $f(\kappa)=h'(\kappa)=0$ and
$\alpha=1$.

Equation (13) describes the self-consistent evolution of a binary
containing extended components due to emission of gravitational
radiation. Our equations differ from those for a point-mass model
in two new functions $f(L)$ and $h(L)$. Precisely these functions
lead the differences in the evolution for these two models. The
most important difference is the deviation of the function
$\alpha(L)$ from unity in the model with extended components. This
function is presented in a form convenient for comparison with
post--Newtonian corrections to the law of motion for two point
masses in a binary system.

The evolution of a binary system with extended components
differs most from the evolution for two point masses in the
final stages, when the effect of tidal distortions becomes
significant. Therefore, below, we will compare two cases: a
binary system consisting of two extended components and
binary consisting of two point masses.

The next set of figures each show two lines: the thick line
corresponds to the case of extended masses ant the thin line to
the point-mass case. In addition, the number of orbital periods
until contact is given on the upper axis on each plot.

Figure 4 presents the right-hand side if differential equation
(13) in the form of convergence velocity of the components
$V=\slantfrac{1}{2}\stackrel{\bullet}{L}$ (in units of velocity of
light $c$) as a function of the distance between the components.
The characteristic orbital velocity of the system
$\slantfrac{1}{2}\Omega L$ is shown by the dashed line. Note that
the case of extended masses yields a higher velocity:  at the end
of convergence, this velocity is $0.016c$ for the case of extended
masses and $0.005c$ for the point-mass case.  Note also that the
convergence velocity at the moment of contact would increase to
$0.114c$ if we did not take into account both the mass
distribution and the `geometrical' distribution of the components;
i.e., if we assumed that stars remain spheres with a fixed radius
$R_1$.

Figure 5 presents the characteristic convergence time of the
components $\tau=L/\stackrel{\bullet}{L}$ in units of orbital
period $P_{orb}$ as a function of the distance between the
components. The dashed line corresponds to the characteristic
hydrodynamic time, which indicates the rate at which equilibrium
is reached inside a star. We can see from this plot that the
star as a whole is nearly in hydrodynamic equilibrium right up
to the moment of contact.

\begin{figure}[t]
\hbox{\psfig{figure=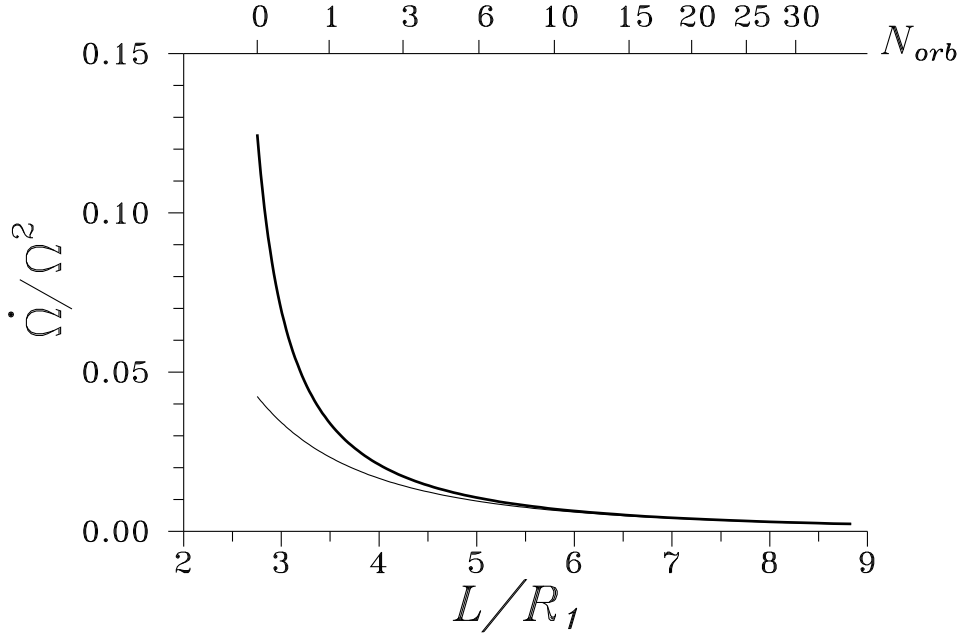,width=5in}}
\caption{\sl Ratio of angular and centrifugal accelerations as a
function of the distance between components for the cases of
extended masses (thick line) and point masses (thin line).}
\end{figure}

A comparison of the angular acceleration
$\stackrel{\bullet}{\Omega} L$ and centrifugal acceleration
$\Omega^2L$ of the system is presented in Fig.~6 in the form of
the ratio of these two quantities. This figure indicates the
errors in our estimates of the power of the gravitational-wave
radiation, which were obtained assuming that the components move
along circular orbits at constant speed.

\begin{figure}[p]
\hbox{\psfig{figure=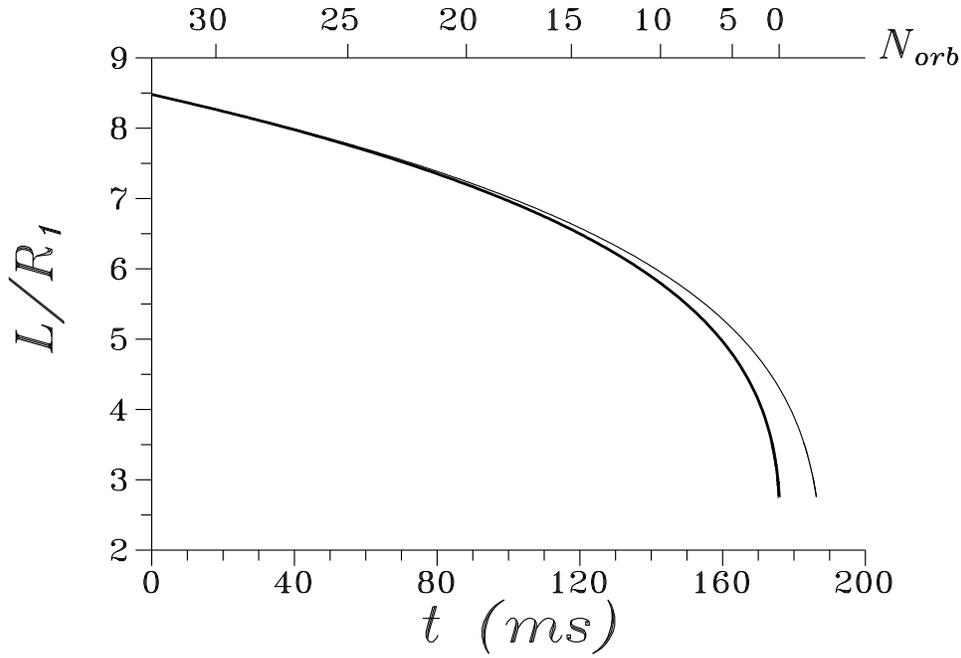,width=5in}}
\caption{\sl Time dependence of the semi-major axis $L(t)$ for
model with extended masses (thick line) and point masses (thin
line).}
\end{figure}

\begin{figure}[p]
\hbox{\psfig{figure=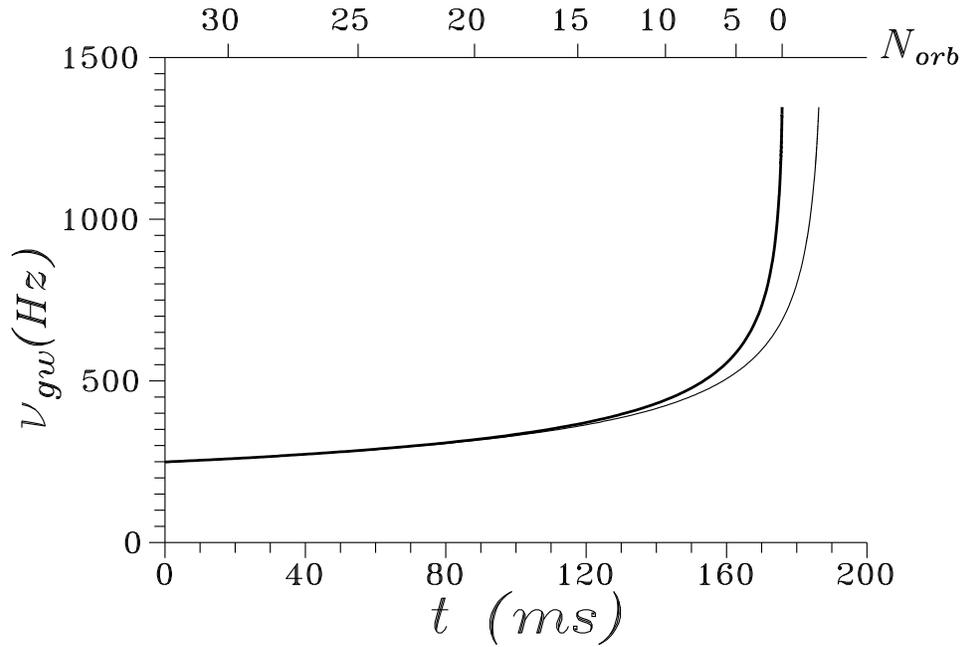,width=5in}}
\caption{\sl Time dependence of the frequency of gravitational
waves (in Hz) for models with extended masses (thick line) and
point masses (thin line).}
\end{figure}

\begin{figure}[t]
\hbox{\psfig{figure=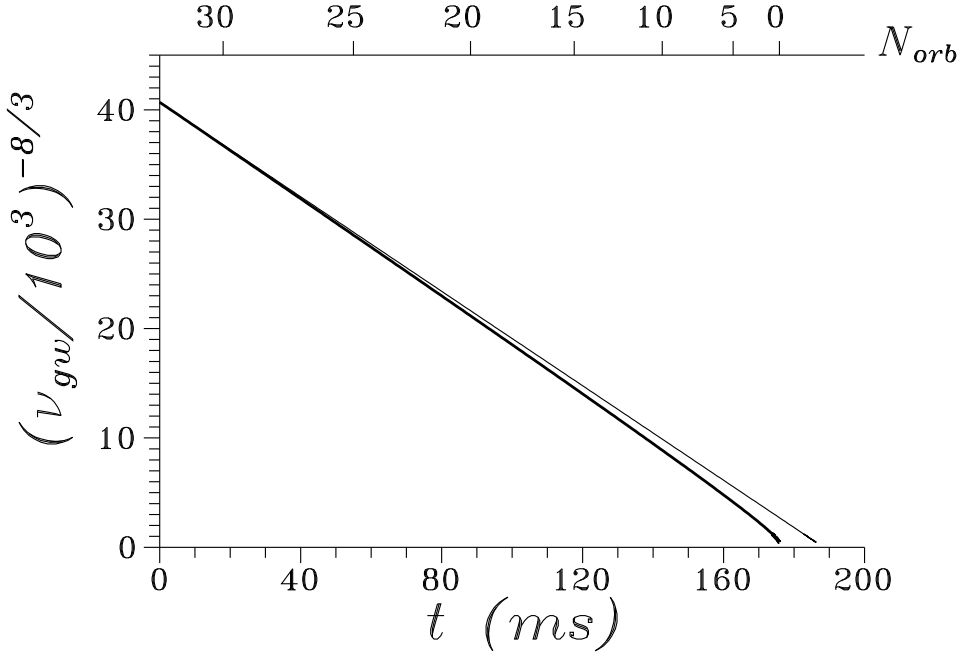,width=5in}}
\caption{\sl Linear gravitational-wave frequency to the
$-\slantfrac{8}{3}$ power for
models with extended masses (thick line) and point
masses (thin line). This quantity is a linear function of time
in the point-mass case. }
\end{figure}

\begin{figure}[p]
\hbox{\psfig{figure=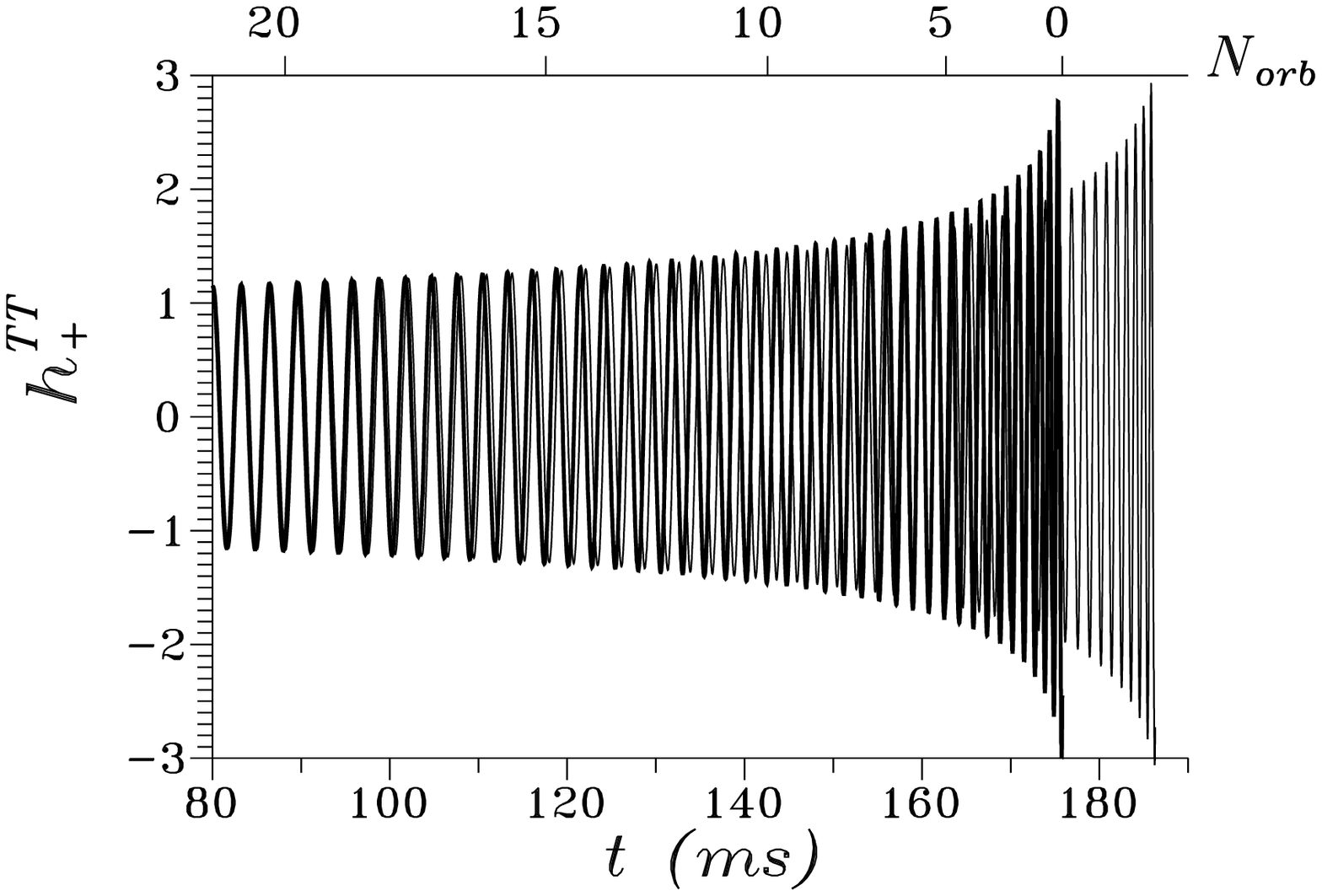,width=5in}}
\caption{\sl Wave forms $h_{+}^{TT}$ for the signals from two
merging neutron stars. We can see that a binary system with
extended masses (thick line) reaches contact 10.5 ms earlier
than the point-mass system (thin line). The phase shift at the
time of contact is $\delta {\cal N}=5.97$.}
\end{figure}

Figures 7--9 depict the solution of the resulting differential
equation for $L$ in the form of the functions $L(t)$,
$\nu_{gw}(t)$ ($\nu_{gw}=2\Omega/(2\pi)$) and
$\nu_{gw}^{-8/3}(t)$. This last relation is interesting in that
it should be linear in the case of point-mass binary, since
$\stackrel{\bullet}{L} \sim L^{-3}$; $L^4 \sim t_0-t$; $\nu_{gw}
\sim L^{-3/2}$).

\begin{figure}[p]
\hbox{\psfig{figure=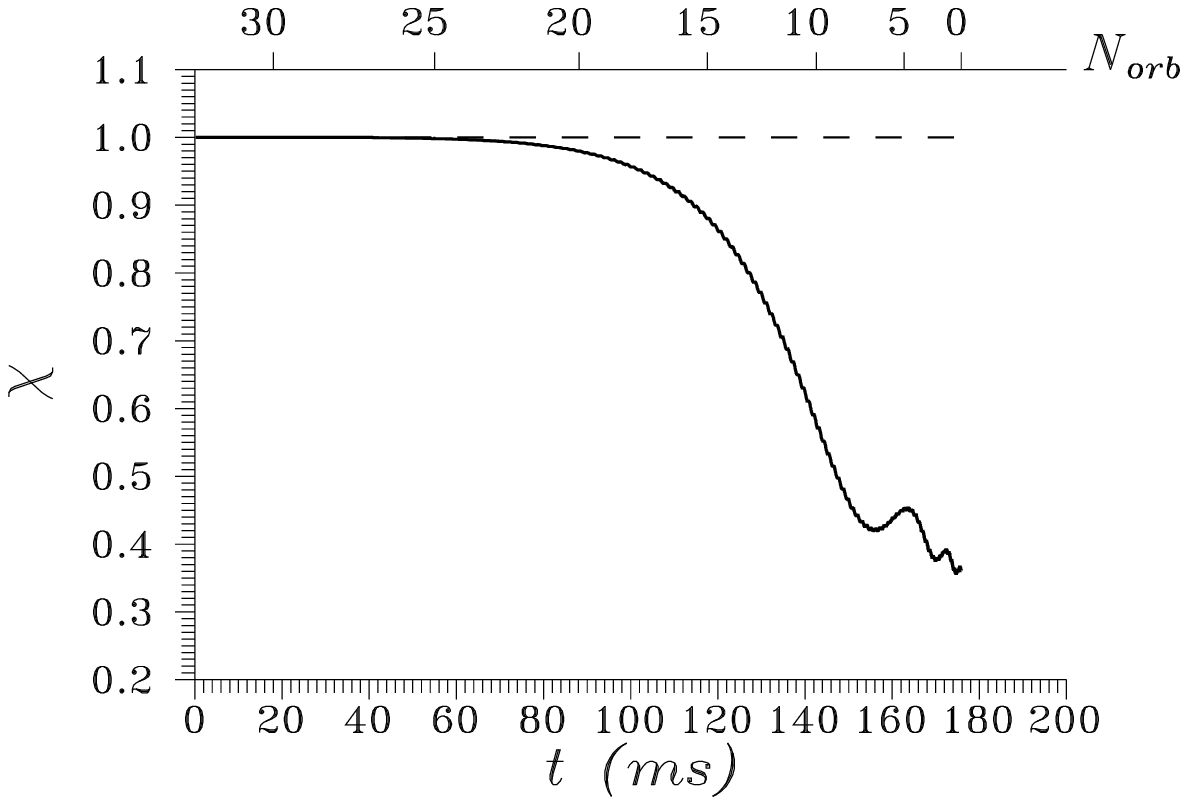,width=5in}}
\caption{\sl Time dependence of the correlation coefficients of
the signals from extended-mass and point-mass systems.}
\end{figure}

\begin{figure}[p]
\hbox{\psfig{figure=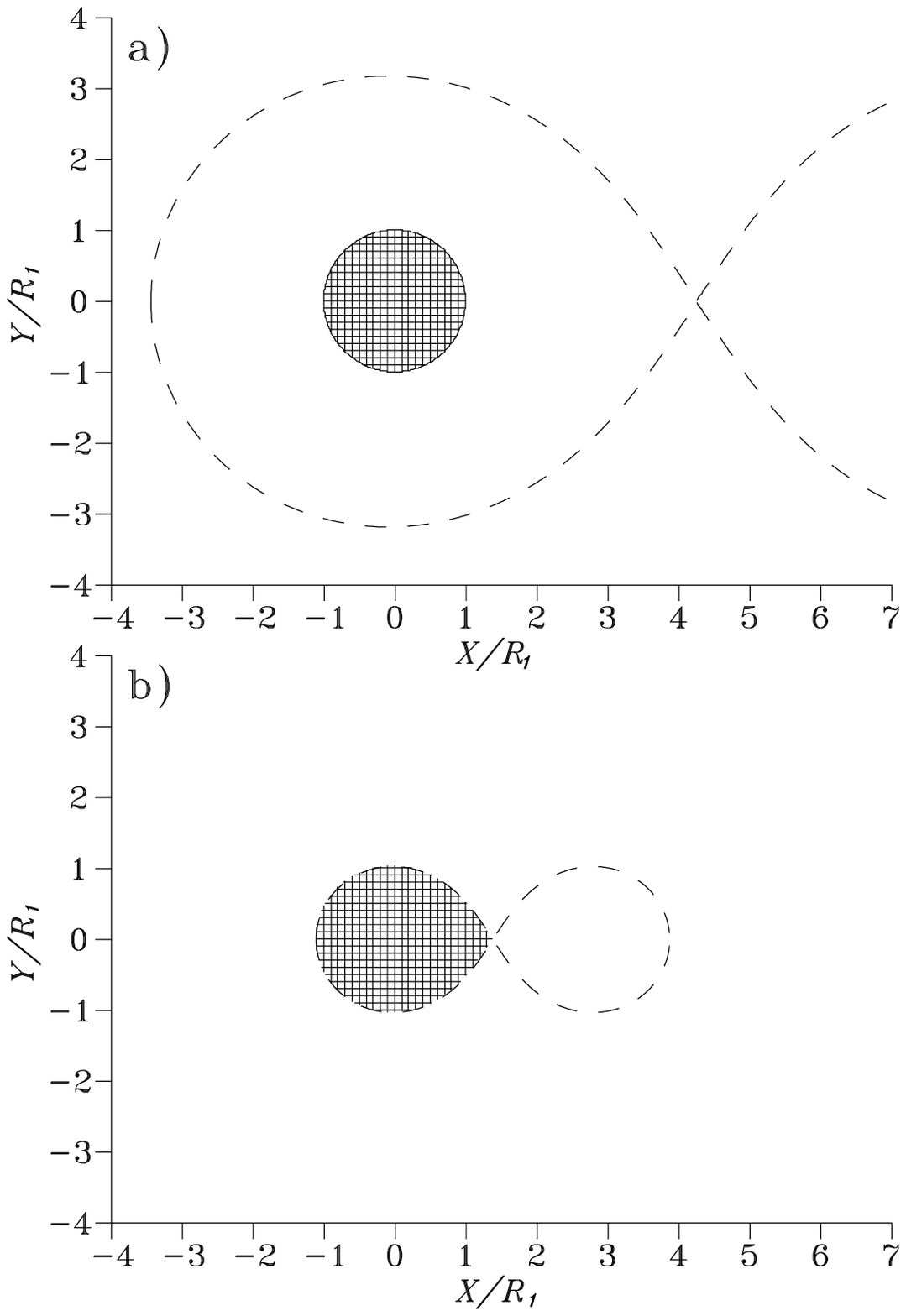,width=6in}}
\caption{\sl Binary system configurations (a) at the beginning of
the calculation ($t=0$, $L=85$~km) and (b) at the
time when the Roche lobes of the components are filled
(contact). The time interval between these configurations id
17.8 ms, which corresponds to 32 revolutions of the system. }
\end{figure}

The waveforms $h_{+}^{TT}$ for the signals from two merging
neutron stars are shown in Fig.~10. We can see that the stars in
a binary with extended masses come into contact 10.5 ms
earlier than in the point-mass model. The phase shift for these
two cases at contact $\delta {\cal N}=5.97$.

The time dependence of the correlation coefficient $\chi$
between the signals from the point-mass and distributed binary
systems is shown in Fig.~11. This correlation coefficient is
determined by the formula

$$
\chi(t)={\int\limits_0^t h_1 h_2\,dt}\left/
{\left(\int\limits_0^t h_1^2\,dt
\int\limits_0^t h_2^2\,dt\right)^{1/2}}\right.\,,
$$
where $h_1$ and $h_2$ are the signals from the extended-mass and
point-mass systems, respectively. The correlation is disrupted
70 ms before contact.

Figure 12 shows the binary system configurations (a) at the
beginning of the calculation ($t=0$, $L=85$~km) and (b) at the
time when the Roche lobes of the components are filled
(contact). The time interval between these configurations id
17.8 ms, which corresponds to 32 revolutions of the system.

Note that the Jeans criterion for the dynamic stability of the
system [10]

$$
J_{rot} < \slantfrac{1}{3} J_{orb}\,,
$$
remains undisrupted right up to the time of contact ($J_{rot}$
is the axial angular momentum of the components, and $J_{orb}$
is their orbital angular momentum).

\section{CONCLUSION}

After gravitational-wave detectors begin to operate, the first
date on the frequency and characteristics of merging binary
neutron stars will become available. The gravitational-wave
signals from such events will be detected by convolving a known
waveform (the calculated signal from two neutron stars in a
point-mass model, for example) with the output of the
gravitational-wave detector. The output of the
gravitational-wave detector can be represented in the form

$$
    s(t)=A\cdot S(t) +n(t) \,,
$$
where $S(t)$ is a given waveform with unknown amplitude and
$n(t)$ is detector noise. The search for a signal using a
least-square method or other suitable technique reduces to
finding the minimum of a functional with respect to the unknown
parameter $A$, the signal amplitude. This procedure can be
rewritten in terms of integrals of the form
$\displaystyle \int^t_0 S^2(t)\,dt$, and so forth.

However, if the convolution is performed using a signal of
the wrong form (in our case, the signal corresponding to two
point masses) the effective signal will be diminished. This may
result in false estimates of the signal level. Figure 11 shows
the convolution of the signal corresponding to our model with
that for a pair of two point masses. We can see that the
convolution with the standard waveform leads to the rapid
decrease of the registered signal $\sim70$~ms before contact.
Note also that the region over which $\alpha(L)$ differs
appreciably from unity falls into the frequency band for the
peak sensitivity of planned gravitational-wave detectors [18].
Therefore, it is important to take into account the distortion
of the waveform when attempting to register a gravitational-wave
signal.

It is evident that the distortions of the waveforms of
gravitational-wave signals will impede detection of
gravitational-wave events. On the other hand, precisely these
distortions may prove to be very important for studies of the
internal structure and equation of state of neutron stars.

\section*{ACKNOWLEDGMENTS}

The authors thank N.I.Shakura, K.A.Postnov, and S.B.Popov for
useful discussions. This work was partially supported by Russian
Foundation for Basic Research (project no. 97-02-16486).

\section*{REFERENCES}

\begin{enumerate}

\item Phinney, E.S. 1991, ApJ, 380, L17

\item Narayan, R., Piran, T., \& Shemi, A. 1991, ApJ, 379, L17

\item Schutz, B.F. 1986, Nature, 323, 310

\item Lipunov, V.M., Nazin, S.N., Panchenko, I.E., Postnov,
K.A., \& Prokhorov, M.E. 1995, A\&A, 298, 677

\item Thorne, K.S. 1987, in {\it Three Hundreds Years of
Gravitation}, eds. S.W.Hawking, W.Israel, Cambrige Univ. Press,
Cambridge, NY, p.330

\item Abramovici, A., et al. 1992, Science, 256, 325

\item Lincoln, C.W., \& Will, C.M. 1990, Phys.Rev.D, 42, 1123

\item Blanchet, L., Damour, T., \& Iyer, B. 1995, Phys. Rev.
Lett., 74, 3515

\item Zakharov, A.V. 1996, Astron. Reports, 40, 552

\item Lai, D., Rasio, F.A., \& Shapiro, S.L. 1993, ApJ, 406, L63

\item Ruffert, M., Rampp, M., \& Janka, H.-Th. 1997, A\&A, 321,
991

\item Kuznetsov, O.A. 1995, Astron. Reports, 39, 450

\item Landau, L.D., \& Lifshitz, E.M. 1975, {\it The classical
theory of fields}, Pergamon, Oxford

\item Chandrasekhar, S. 1939, {\it Introduction to the study of
stellar structure}, Univ. Chicago Press, Chicago

\item Zel'dovich, Ya.B., \& Novikov, I.D. 1967, {\it
Relativistic Astrophysics}, Nauka, Moscow (in Russian)

\item Landau, L.D., \& Lifshitz, E.M. 1980, {\it Statistical
physics. Part I}, Pergamon, Oxford

\item Brown, G.E., Weingartner, J.C., \& Wijers, R. 1996, ApJ,
463, 297

\item Thorne, K.S. 1993, in {\it Particle Astrophysics}, eds.
G.Fontaine, J.Tr\^an Thanh V\^an,  Editions Frontieres,
Gif-sur-Yvette, p.375

\end{enumerate}

\end{document}